\documentclass{vldb}
\usepackage{graphicx}
\usepackage{balance}  
\usepackage{CJKutf8} 

\usepackage{booktabs}
\makeatother \makeatletter
\newif\if@restonecol
\makeatother

\usepackage[linesnumbered,ruled,vlined]{algorithm2e}

\usepackage{color}
\usepackage{colortbl}
\usepackage{url}
\usepackage{verbatim}
\usepackage[shortlabels]{enumitem}
\usepackage{amssymb}
\usepackage{multirow}
\usepackage{lscape}
\usepackage{float}
\usepackage[T1]{fontenc}
\usepackage{tgtermes}
\usepackage{listings}
\usepackage{lipsum}
\usepackage{mwe}
\usepackage{subfig}
\usepackage{authblk}
\usepackage{url}
\usepackage{soul}
\usepackage{color}

\setlist[itemize]{leftmargin=*}
\setlist[enumerate]{leftmargin=*}

\usepackage{epstopdf}

\definecolor{mygray}{gray}{.9}
\definecolor{mygreen}{rgb}{0.0, 0.5, 0.0}
\definecolor{myred}{rgb}{0.8, 0.0, 0.0}
\definecolor{mygreen1}{rgb}{0.03, 0.91, 0.87}
\definecolor{mypink}{rgb}{1.0, 0.0, 0.5}
\definecolor{mycorn}{rgb}{0.98, 0.93, 0.36}

\newcommand{\yuliang}[1]{{\it\small\textcolor{blue}{[[[ {#1}\ --yuliang ]]]}}}
\newcommand{\wctan}[1]{{\it\small\textcolor{red}{[[[ {#1}\ --wangchiew ]]]}}}
\newcommand{\aaron}[1]{{\it\small\textcolor{red}{[[[ {#1}\ --aaron ]]]}}}

\newcommand{\hide}[1]{{\it\small\textcolor{blue}{[[[ {#1}\ --hide ]]]}}}

\usepackage{dsfont}

\newcommand{\system}{{\sc OpineDB}}

\vldbTitle{Towards Productionizing Subjective Search Systems}
\vldbAuthors{A. Feng, S. Chen, Y. Li, H. Matsuda, H. Tamaki, and W.-C. Tan}
\vldbDOI{ }
\vldbVolume{13}
\vldbNumber{12}
\vldbYear{2020}

\begin{document} 

\title{Towards Productionizing Subjective Search Systems}



\numberofauthors{1} 


\renewcommand\Authfont{\fontsize{11}{12}\selectfont}
\author[1,2]{Aaron Feng}
\author[1]{Shuwei Chen}
\author[1]{Yuliang Li}
\author[1,2]{Hiroshi Matsuda}
\author[1,2]{Hidekazu Tamaki}
\author[1]{Wang-Chiew Tan}

{
\makeatletter
\renewcommand\AB@affilsepx{\qquad\protect\Affilfont}
\affil[1]{Megagon Labs}
\affil[2]{Recruit Co.,Ltd.}
\makeatother
}
{
\makeatletter
\renewcommand\AB@affilsep{\protect\\\protect\Affilfont}
\renewcommand\AB@affilsepx{\protect\\\protect\Affilfont}
\affil[ ]{\{aaron, vincent, yuliang, hiroshi\_matsuda, hide\_tamaki, wangchiew\}@megagon.ai}
\makeatother
}

\renewcommand{\hl}[1]{#1}
\renewcommand{\aaron}[1]{}
\renewcommand{\yuliang}[1]{}
\renewcommand{\wctan}[1]{}
\renewcommand{\hide}[1]{}

\maketitle

\begin{abstract}
Existing e-commerce search engines typically support  search only over objective attributes,
such as price and locations, leaving the more desirable subjective attributes, such as romantic vibe and work-life balance unsearchable. 
We found that this is also the case for Recruit Group,
which operates a wide range of online booking and search services, including jobs, travel, housing, bridal, dining, beauty, and where each service is among the biggest in Japan, if not internationally.

In this paper, we present our progress towards productionizing a recent subjective search prototype (OpineDB) developed by 
Megagon Labs for Recruit Group.
Several components within OpineDB are enhanced to satisfy production demands, 
including adding a BERT language model pre-trained on massive hospitality domain review corpora.

We also found that the challenges of productionizing the system are beyond enhancing the components. 
In particular, an important requirement in production quality systems is to instrument a proper way of 
measuring the search quality, which is extremely tricky when the results of such search systems are subjective. 
This led to the creation of a high-quality benchmark dataset from scratch, 
involving over 600 queries by user interviews and a collection of more than 120,000 query-entity relevancy labels.
Also, we found that the existing search algorithms do not meet the search quality standard required by production systems. 
Consequently, we enhanced the ranking model by fine-tuning several search algorithms 
and combining them under a learning-to-rank framework.
The model achieves 5\%-10\% overall precision improvement
and 90+\% precision on more than half of the benchmark testing queries
making these queries ready for AB-testing. 
While some enhancements to OpineDB can be immediately applied to other verticals,
our experience so far reveals that obtaining the benchmark dataset and fine-tuning ranking algorithms are specific to each domain and cannot be avoided.
\end{abstract}

\newpage
\section{Introduction}\label{sec:intro}

A subjective search system finds entities that satisfy subjective 
requests, such as ``{\sl hotels with clean rooms and close to really good cafes}''.
Recently, a crowdsourcing task in  \cite{opinedb} revealed that users' search criteria are largely subjective;
across 7 common verticals, 
over 50\% and up to 82\% of users' most frequent search criteria are subjective.
Existing e-commerce search engines, however, typically
allow users to search only over the objective attributes, such as price, rating, or location, and may further allow the results to be filtered over a predefined set of subjective attributes.

\begin{figure}[!ht]
    \centering
    \includegraphics[width=0.48\textwidth]{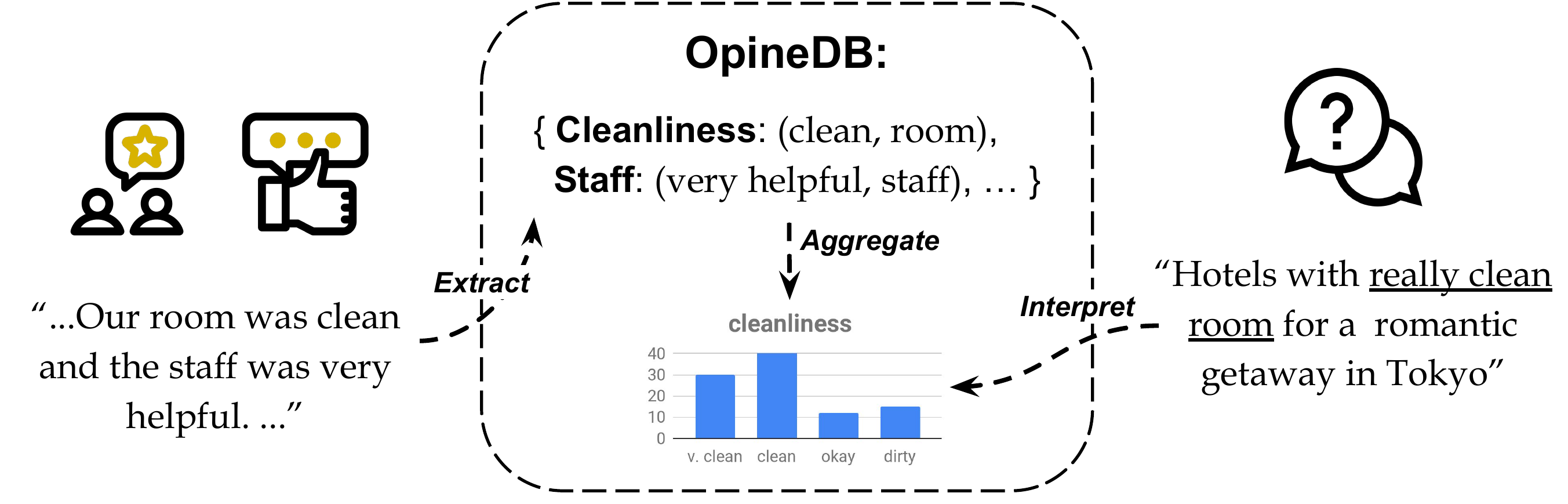}
    \caption{\small Overview of \system. Given the subjective query, \system\ interprets the query into 
    one of the pre-defined subjective attributes (i.e., ``\textbf{cleanliness}'' in this example) and
    ranks the entities using features (i.e., a histogram of the different levels of cleanliness) 
    summarized from opinions extracted from reviews. }
    \label{fig:opinedb}
\end{figure}

We conducted a similar survey for the hospitality domain in the Japanese market and obtained similar results. With these encouraging results, we proceeded to develop a production-quality subjective search system for the hospitality domain.
We started with \system~\cite{opinedb, voyageur}, a subjective search prototype that is developed by Megagon Labs, a research arm of Recruit Group. OpineDB extracts subjective data (e.g., cleanliness of rooms, friendliness of staff etc.)
from review text, explicitly 
models subjective attributes in a schema (e.g., a set of attributes such as cleanliness, staff, breakfast, value for money etc.), 
and supports free-text querying
over the aggregated subjective data (e.g., ``{\sl find me hotels with really clean rooms for a romantic getaway}'').
See Figure \ref{fig:opinedb} for an overview. 

It turns out that taking \system\ to a production-level a search system is more challenging than expected, as we shall describe next.

\smallskip
\noindent
\textbf{Matching.} \yuliang{TODO}
The quality of \system's search results is sensitive to its
underlying subjective schema.
The subjective schema is constructed from reviews and contains the most important subjective attributes that represent what people search for in the target domain.
\system{} generates query results by first matching the request
against the schema attributes. It also approximates the attributes in the schema to match the request in case a direct match cannot be determined.
At the core of the matching process is a phrase2vec \cite{word2vec} embedding module which is used by 
\system\ to map the text of the queries to one of the attributes (e.g., ``really clean room'' $\rightarrow$ 
\texttt{cleanliness}). {\em We found that results of requests that can be matched well
with the schema's attributes
are of significantly higher quality 
than those that are not directly captured by the schema.}
To address the challenge of improving the quality of query results obtained when requests do not match well with the schema attributes, 
we leverage the BERT~\cite{bert} language model which was recently shown to improve the
already powerful Google search engine~\cite{bert-google}. In particular, the BERT model
pre-trained on a large hotel review corpus
can retrieve entities based on textual similarity between queries and review text
without relying on the subjective schema~\cite{rte}.
As we discuss in Section \ref{sec:algo}, the BERT-based method complements \system\ and
combining them under a {\em learning-to-rank} framework yields the best overall search quality.

\smallskip
\noindent
\textbf{Benchmarking. }
The second major challenge was that there does not exist any benchmark dataset for evaluating subjective search systems. 
Even worse, we could not obtain subjective queries from Recruit Group's online services; their systems capture only objective searches.
Hence, we found it necessary to create a benchmark
of carefully curated queries that can
represent the incoming real query distributions. 
For each query in the benchmark, 
we will also need to construct the set of entities that satisfies the subjective aspects of the query.
As this task is crowdsourced and our labeling task is subjective by nature, quality control is of utmost importance. 
Finally, it is expensive to determine the set of entities for each query by labeling all the query-entity pairs.
So we need to determine a selected subset of ``most valuable'' pairs to send
to the crowd-workers.
We describe in Section \ref{sec:dataset} our nontrivial crowdsourcing process to 
construct this benchmark.
Currently, our benchmark consists of over (1) 600 subjective queries from crowd-workers
and (2) 40,000 query-entity relevancy labels. This benchmark has enabled us to systematically
evaluate different search algorithms with confidence.

\smallskip
\noindent
{\bf Localization. } 
Building the subjective search system requires significant engineering overhaul to 
provide Japanese language support since Recruit Group consists of many Japanese online services.
To adapt \system{} for the Japanese language, we
(1) implemented GiNZA~\cite{ginza}, a spaCy-like Japanese NLP library, to deploy \system{} on Japanese reviews/queries and 
(2) pre-trained the BERT models on a large corpus of Japanese review text~\cite{rte}. 

\smallskip
In what follows, we focus our discussions on the main efforts, which are the construction of the benchmark and the evaluation of the enhanced \system{}. 
Section \ref{sec:dataset} describes our
crowdsourcing process for constructing the evaluation benchmark.
Section \ref{sec:algo} introduces our learning-to-rank framework for combining
\system\ with BERT. We experimentally validated the performance of our enhanced 
\system{} in Section \ref{sec:experiments} and conclude in Section \ref{sec:conclusion}.
\section{Benchmark Dataset}\label{sec:dataset}

In this section, we describe how we create the benchmark dataset
for evaluating subjective search systems by crowdsourcing.
Since we targeted a Japanese hospitality platform,
the benchmark dataset consists of two parts: (1) a set of subjective queries for hotel and 
(2) a label set, i.e., the query-hotel relevancy labels depicting how well the hotel satisfies the query. 

\subsection{Subjective Queries}

To ensure that the benchmark queries are meaningful in the real production setting, we need to ensure that
the queries are representative of 
real subjective requests. To do this,
\hl{we first collected around 900 hotel reservation dialogues from pairs of workers, where one plays the role of a hotel reservation agent and the other plays the role of a customer seeking to reserve a hotel.}
Additionally, we interviewed over 500 Japanese crowd-workers and asked them to describe their requests when making hotel reservations.
By analyzing these dialogues and interview results, we found that the requests always contain
an area that specifies their desired travel destination.  
An area can be a specific address, a district, or a region.
In fact, most Japanese hotel booking applications require customers to specify a location to begin their search.

These user studies eventually yielded 336 queries, where each query is of the form ``(\texttt{query\_text}, \texttt{area})''. For example, (hotels with beautiful sea view, Atami).
The areas range from small towns such as Ueno
to large metropolitan areas that contain multiple prefectures such as the Kanto region.

\hl{To collect even more queries, we crowdsourced again and asked workers to write a tagline summary sentence of Japanese hotel reviews posted on \emph{Jalan.net}, which is one of the largest online hotel booking sites in Japan.}
For each tagline summary, we specify the prefecture of the hotel which the tagline summary is associated with.
We collected another 328 queries this way.
The statistics of the subjective query set are shown in Table \ref{tab:queries}.

\setlength{\tabcolsep}{4pt}
\begin{table}[ht!]
    \small
    \centering
    \vspace{-2mm}
    \caption{\small{Summary Statistics of Query Set.} }\label{tab:queries}
\begin{tabular}{ccccc}
\toprule
  Total \# of    &  Total \# of  &  \multicolumn{3}{c}{Query Length} \\ 
   Queries &  Distinct Areas & Average & Shortest & Longest \\ \midrule
  664   & 79 & 16.03 & 4 & 84  \\ \bottomrule
\end{tabular}
\end{table}

\subsection{Relevancy Labels}

The benchmark dataset also contains a set of query-hotel relevancy labels which indicates whether the hotel is relevant for the associated query.
\hl{For each query-hotel pair, we ask crowd-workers to examine the hotel front page in Jalan.net, including photos, descriptive text, reviews, etc., and decide whether the hotel is relevant to the query within 90 seconds.}
However, as this task is inherently subjective and there are many query-hotel pairs,  
we need to ensure (a) quality control over the labels obtained, and (b) prioritize query-hotel candidates for labeling.

\smallskip
\noindent
{\bf Resolving Subjectivity.}
The relevancy labeling task is highly subjective.
For example, a customer may consider one hotel as clean but another customer may consider the same hotel as dirty.
Thus, there may be disagreements between different crowd-workers.
To make matters worse, workers may make a decision based on different pieces of information (e.g., pictures of hotel rooms or reviews or descriptions).

To improve the quality of the labels we obtain, we instruct the crowd-workers to make each judgement only after they have examined
all the information: the hotel description, photos, and hotel reviews. To avoid malicious or random labeling, we require each worker to justify their labels with evidence which will be verified later.
We further implemented a majority voting to ensure that we pick only labels where there is a certain degree of agreement among workers.
Each query-hotel pair was labeled by 3 unique workers, and a pair is relevant only if at least 2 out of 3 workers consider that the hotel is relevant to the query. 




\smallskip
\noindent
{\bf Iterative Labeling. }
Since there is a large number of candidate pairs, it is prohibitively expensive to label all of them.
To reduce the labeling cost, we label only entities (i.e., hotels) that are returned by the search algorithms and improve the labeling {\em iteratively}.
By labeling iteratively, 
we can adapt to the ``importance'' of each query-hotel pair, which can change over time as we develop the search system.
For example, for the precision@10 metric (i.e., precision for the top-10 entities), we first label
only the top-10 entities returned by each algorithm.
At the very beginning, the most important pairs were the ones returned by the first version of \system.
It is also important to understand how much \system\ is better than a baseline algorithm. 
We considered random ordering as the first baseline and labeled its resulting pairs as well.
In the next iteration, as we fine-tuned the schema of \system\ and added more models (BERT and BM25),
we also labeled their top-10 entities to evaluate their results. The pairs already labeled in the previous round are not re-labeled.
Note that the decision of fine-tuning and adding new models 
were based on our error analysis on the results of \system. This means that we did not know the importance of those candidates ahead of time.
We repeated this process 3 times to obtain the final dataset and models.

Eventually, we collected 40,886 unique query-hotel pair labels based on 122,658 judgments of crowd-workers. 
Among all unique pairs, 23,862 pairs are labeled as relevant. We summarize the statistics of these labels in Table \ref{tab:labels}.

\setlength{\tabcolsep}{2.5pt}
\begin{table}[!ht]
\small
    \centering
\vspace{-2mm}
    \caption{\small{Summary of the Relevancy Label Dataset. See Section \ref{sec:algo} and \ref{sec:experiments} 
    for the details of the tested models. We underlined the model that produces the best performance at each round.}}\label{tab:labels}
\resizebox{\columnwidth}{!}{%
\begin{tabular}{cccccc}
\toprule
\textbf{Rounds} & \textbf{\#Queries} & \textbf{\#Pairs} & \textbf{\%Pos.} & \textbf{Prec@10} & \textbf{New Tested Models}                                \\ \midrule
1      & 500       & 6,000    & 52.6\%    & 0.622        & \underline{OpineDB}, Random                              \\
2    & 500       & 28,386   & 60.3\%    & 0.662       & \underline{OpineDB+}, BERT, BM25                       \\
3  & 664       & 40,886   & 58.4\%    & 0.755       & Rating, \underline{Logit}, LambdaMart \\ \bottomrule
\end{tabular}}

\end{table}


\section{Ranking Algorithms}\label{sec:algo}

In this section, we briefly summarize the ranking algorithms that we implemented
and experimented with.
Our implementation started with \system\ as a single ranking algorithm.
Intuitively,
\system\ answers subjective queries by mapping the query term to one or more \emph{subjective attributes}.
Each attribute captures  an important aspect of the underlying subjective data (e.g., hotel reviews).
\system\ models each attribute as a \emph{linguistic domain} consisting of all linguistic variations
that describe the attribute. For example,
\begin{align*}
& \textbf{cleanliness: } [\text{clean room}, \text{very clean hotel}, \text{filthy carpet} \dots ] \\
& \textbf{staff: } [\text{friendly staff}, \text{exceptional service}, \text{warm welcome} \dots ]
\end{align*}

In \cite{opinedb}, \system{} extracts these phrases from the review corpus and summarizes them using an opinion extraction pipeline. 
Although this pipeline exists for English reviews,
building one for Japanese is not trivial due to the lack of mature NLP tools and resources for Japanese. 
To overcome this challenge, we implemented and open-sourced GiNZA \cite{ginza}, a Japanese NLP library, for building an extractor based on dependency parsing and pattern matching.

Not surprisingly, the search quality of \system\ heavily depends on the set of subjective attributes
and linguistic domains. The first version of \system\ achieved a precision@10 of only 62.2\% (Table \ref{tab:labels}),
but after we fine-tuned the attribute schema on a development set by adding more attributes and phrases,
the precision@10 went up to 66.2\% (denoted as \system+, row 2 of Table \ref{tab:labels}).

While \system{} uses word2vec~\cite{word2vec} to increase the size of linguistic domains
and fine-tuning the schema can further improve the search quality, 
we notice in our error analysis that it is hard for \system{} when there are queries of meaning 
beyond the combination of words.
By this observation, we implemented a similarity search algorithm by 
constructing sentence embeddings using a BERT model.
\hl{In this method, we train SentencePiece~\cite{kudo-richardson-2018-sentencepiece} and BERT models on review text with over 20 million sentences \cite{rte} and fine-tune on about 300k thousand review-reply sentence pairs.}
For instance, \system\ always returns relevant entities for queries that are covered by the schema like 
\begin{CJK*}{UTF8}{min}``静かな宿'' \end{CJK*}
(quiet place), but it does not perform well in the cases like 
\begin{CJK*}{UTF8}{min}``最低限度のサービスで好きにさせてくれる''\end{CJK*}
(minimum service that let me have my own way).
For the latter query, \system{} wrongly matches hotels with reviews of the meaning close to "minimum service", while the BERT-based approach matches semantically similar sentences like 
\begin{CJK*}{UTF8}{min}`` いい意味でほったらかし''\end{CJK*}
(in a good sense that they leave me alone).

\yuliang{I rephrased the first sentence, pls check.}
With a similar motivation to BERT, we also implemented the following two search algorithms:
\vspace{-2mm}
\begin{itemize}\parskip=0pt
\item \textit{Rating:} 
\hl{This method leverages the structured information like hotel ratings and fine-grained aspect ratings available in the hotel booking platform.}
We compute a weighted sum of the ratings by a word2vec similarity between the subjective query with each aspect. 
\item \textit{Okapi BM25:} BM25 is a variant of the classic TF-IDF ranking algorithm~\cite{ir} 
used in many popular document retrieval systems like \texttt{elasticsearch}.
\end{itemize}

Our experiments reveal that although these algorithms do not out-perform \system{} individually, they complement \system\ as they tend to outperform \system{} for cases where \system{} does not perform well as shown in Table \ref{tab:orth}.
This led to the idea of using learning-to-rank~\cite{liu2009learning} for combining these different ranking models to achieve an improved search quality.

\setlength{\tabcolsep}{4pt}
\begin{table}[!ht]
\small
    \centering
    \caption{\small{Orthogonality with \system{}. Out of the 63 queries that \system{} has precision@10 < 0.5, we show the number of overlapped queries that each ranking algorithm performs better (precision@10 $\geq$ 0.5)}}
    \label{tab:orth}
    \begin{tabular}{ccccc}
    \toprule
    & \textbf{OpineDB} & \textbf{BERT} & \textbf{Rating} & \textbf{BM25} \\
    & Precision@10 $<$ 0.5 & \multicolumn{3}{c}{Precision@10 $\geq$ 0.5} \\
    \midrule
    \textbf{\#queries} & 63 & 25 & 27 & 18 \\
    \bottomrule
    \end{tabular}
\end{table}

Intuitively, learning-to-rank formulates the ranking problem as a supervised learning problem.
To combine the base ranking algorithms (\system{}, BERT, BM25, and Rating),
we featurize each query-entity pair by computing the ranking scores returned by each algorithm.
We then train a learning-to-rank model using these features on a training set extracted from the benchmark dataset.

Specifically, we consider two learning-to-rank models, \emph{Logit} and \emph{LambdaMART}. 
Logit trains a Logistic Regression binary classifier on the binary relevancy labels and,
at prediction time, uses the classifier's probabilistic output as the ranking score.
LambdaMART~\cite{lambdamart} is a popular learning-to-rank algorithm with a similar supervised learning setting
but with the goal of minimizing the average number of inversions in rankings.

\newcommand{\sat}{\mathsf{sat}}
\newcommand{\quality}{\mathsf{quality}}
\newcommand{\satmax}{\mathsf{sat\text{-}max}}

\section{Experiments}\label{sec:experiments}

We evaluate the search quality of the different search algorithms in this section.
Our first result shows that combining the base ranking algorithms (\system, BERT, BM25, and Rating)
using learning-to-rank significantly improves the overall search quality.
Second, we evaluate the algorithms on two special requirements important to our production setting: 
(1) top-k sensitivity and
(2) precision distribution over the benchmark queries.
Top-k sensitivity is important because for hotel booking, the ranked list only contains hotels with available rooms
so the system is expected to high-quality top-k results across different k's.
The query precision distribution is important because when introducing subjective search as new functionality,
we need to identify a subset of highly accurate queries for AB-testing 
so that the online platform can later be confident to fully deploy the system.
Our results show that the current system satisfies both requirements and is ready for the next-stage evaluation.




\subsection{Experiment Setup}

To measure the goodness of a given ranking, we use two very popular metrics, 
(1) Precision@K and (2) Normalized Discounted Cumulative Gain (NDCG)~\cite{ir}.
Precision@K is defined as $M/K$ if $M$ out of the top-$K$ results are labeled as relevant.
Besides the number of relevant results in top-$K$, NDCG also quantifies whether 
the relevant results are ranked higher than the irrelevant results, 
by penalizing highly-ranked irrelevant results.

For the learning-to-rank models, we divide the benchmark dataset by query 
and use 250 queries that occur most often in the hotel reservation dialogues for testing and 
the rest 414 queries for training.
The 250 testing queries are associated with 17,381 unique query-hotel labels, 
and the training set contains 414 queries and 23,505 unique query-hotel labels.
For the fine-tuning of \system{} schema, we also use the same queries in the training set.

We implemented all the ranking models in Python and use libraries shown in Table \ref{tab:library}.
All the experiments are executed using an AWS c5.9xlarge server with 72GB RAM.

\setlength{\tabcolsep}{3.5pt}
\begin{table}[!ht]
\small
    \centering
    \caption{\small{Python Libraries for Ranking Models}}
    \label{tab:library}
    \begin{tabular}{cccccc}
    \toprule
    \textbf{OpineDB} & \textbf{BERT} & \textbf{Rating} & \textbf{BM25} & \textbf{Logit} & \textbf{LambdaMART} \\
    \midrule
    GiNZA & Faiss~\cite{faiss} & Gensim & Elasticsearch & Sklearn & pyltr\\
    \bottomrule
    \end{tabular}
\end{table}

\subsection{Overall Ranking Precision}
Table \ref{tab:precision} shows the precision@K and NDCG@K of 6 ranking algorithms: 
\system, BM25, BERT, Rating, Logit, and LambdaMART already described in Section \ref{sec:algo}.
We choose $K = 10$ and $3$ because top-10 is commonly used in web-based search interface and 
top-3 is suitable for applications such as chatbots and mobile apps.

The results show that \system\ has the highest quality among the 4 base models and
by combining with the other models under learning-to-rank, the accuracy is significantly improved on all the 4 metrics.
Overall, the simple Logistic Regression method (Logit) achieves the best scores and improves over 
\system\ by about 0.09 on Top-10 precision and NDCG, about 0.08 on Top-3 scores.

\setlength{\tabcolsep}{3.5pt}
\begin{table}[!ht]
    \caption{\small{Precisions of Ranking Algorithms.}}
    \label{tab:precision}
    \small
    \centering
    \begin{tabular}{c|c|c|c|c}
    \toprule
    & Prec.@10 & Prec.@3 & NDCG@10 & NDCG@3\\
    \midrule
    \system\ &
    0.664&
    0.693&
    0.648&
    0.691\\
    BERT&
    0.578&
    0.598&
    0.562&
    0.591\\
    Rating&
    0.581&
    0.560&
    0.578&
    0.557\\
    BM25 & 0.657 & 0.667 & 0.661 & 0.666\\
    Logit & {\bf 0.755} & {\bf 0.774} & {\bf 0.742} & {\bf 0.769}\\
    LambdaMART& 
    0.741&
    0.768& 
    0.722&
    0.763\\
    \bottomrule
    \end{tabular}
\end{table}

\subsection{Sensitivity Study on Top-K}
In a production-level search system, the search results are often presented to users 
and usually followed by the user making a hotel or restaurant reservation.
Since the user interface only shows entities that are not fully booked,
we need to ensure that the top-$K$ search results are of high quality across multiple $K$.
This motivates us to verify whether the precision drops significantly when $K$ increases.

\begin{figure}[t]
\centering
  \centering
  \includegraphics[width=\linewidth]{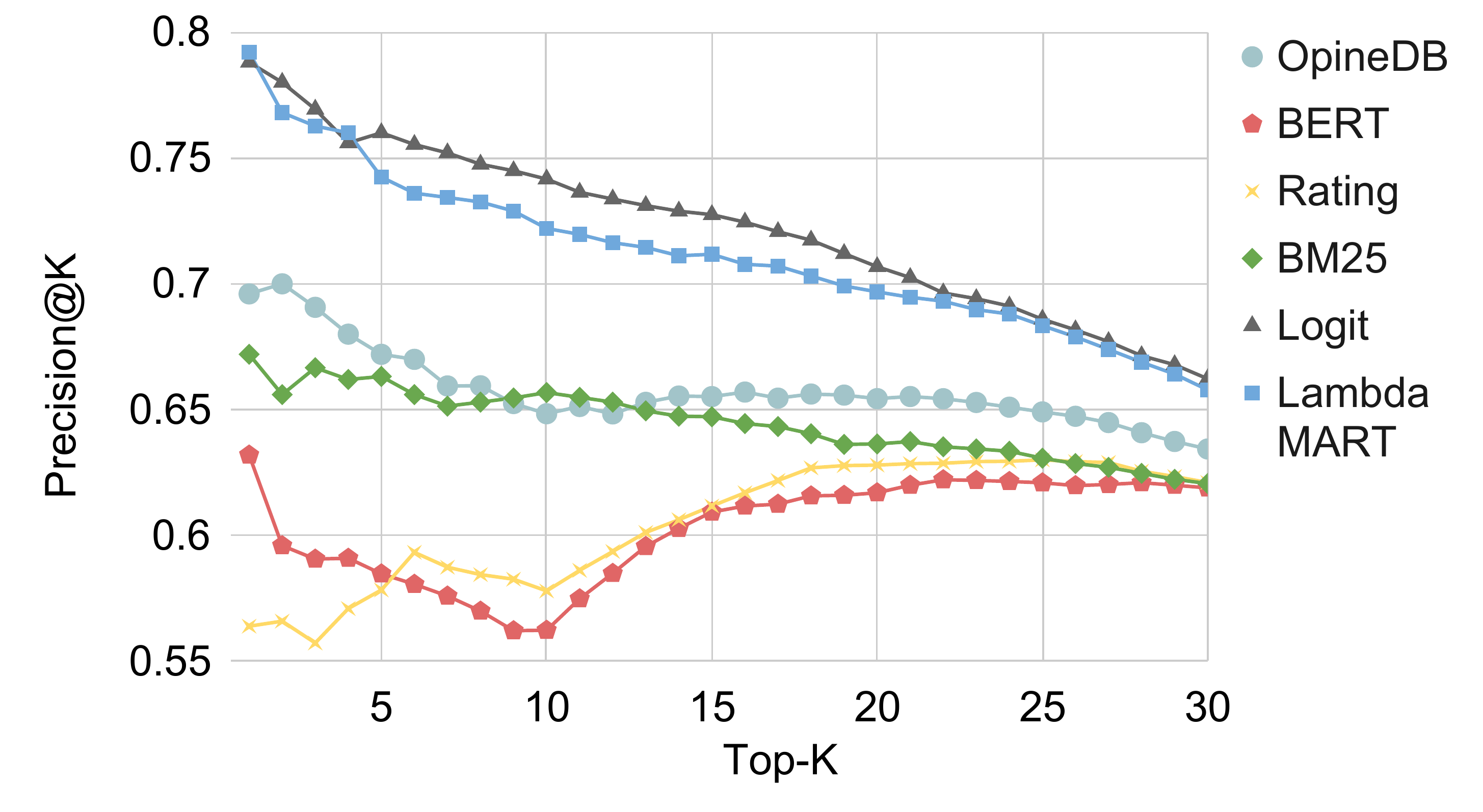}
  \caption{\small{Top-K Sensitivity}}
  \label{fig:sensitivity}
\end{figure}

In Figure \ref{fig:sensitivity}, we can see the precision of the learning-to-rank models drop faster than other baselines.
This is due to our training labels are collected for Top 10 candidates only, but we still see the learning-to-rank models maintain a significant improvement through $K$ from 1 to 30. Even at $K=20$ (i.e., all the top-10 hotels are fully booked), 
the precision of the search results remains 70\%.

\subsection{Precision Distribution by Query}
In a production software setting, there are often many reasons to be careful about introducing new functionality.
One way to get the system being used is to start with some small workload set.
In the subjective search case, this means we can provide a plus to an existing search system with support on subjective queries by simply rerouting some queries to using our approach.
This motivates us to verify how many queries that our approach is performing very well, e.g., 9 out of Top-10 are correct.

\begin{figure}
    \centering
    \subfloat[Precision@10]{\includegraphics[scale=0.26]{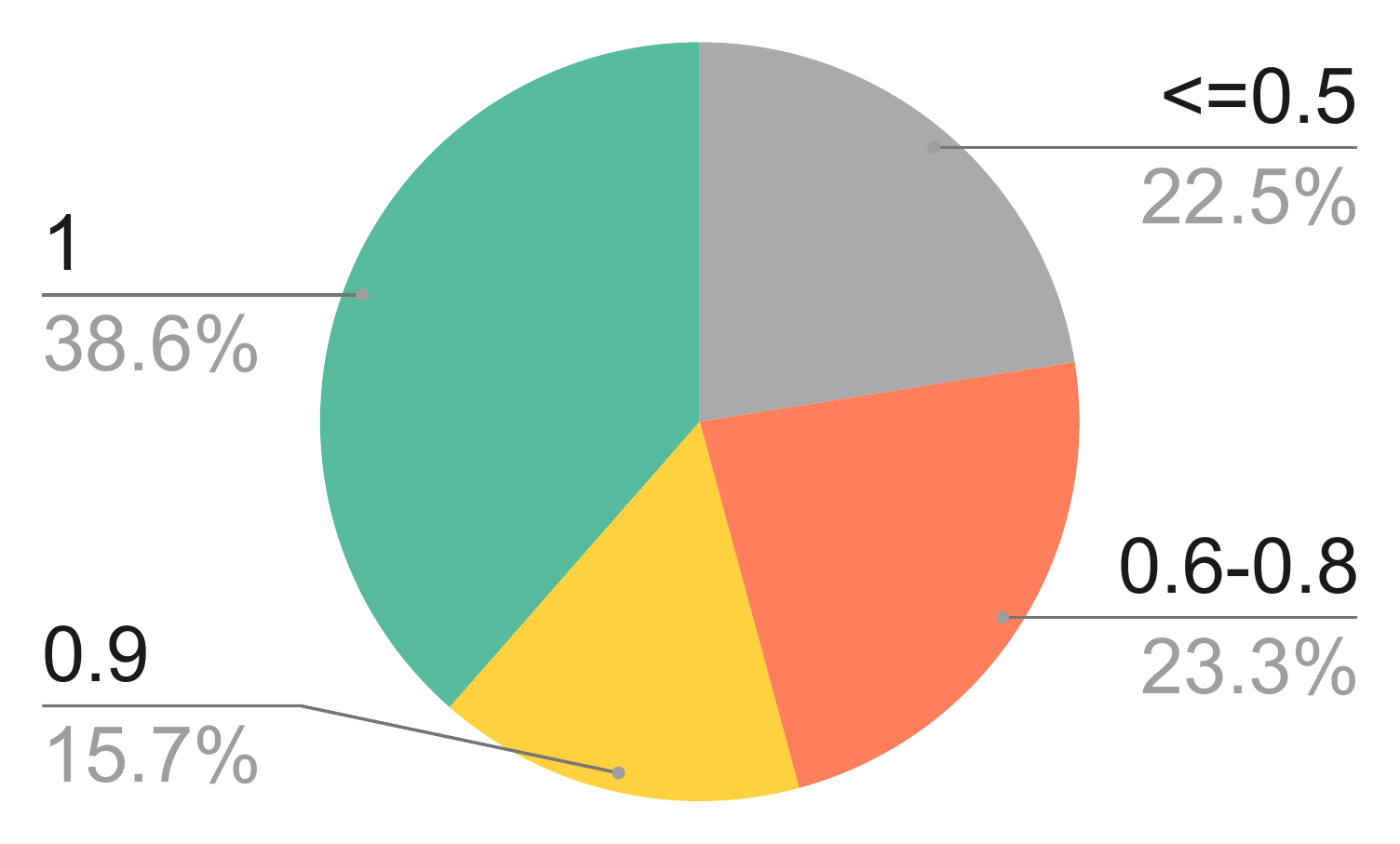}}%
    \qquad
    \subfloat[Precision@3]{\includegraphics[scale=0.27]{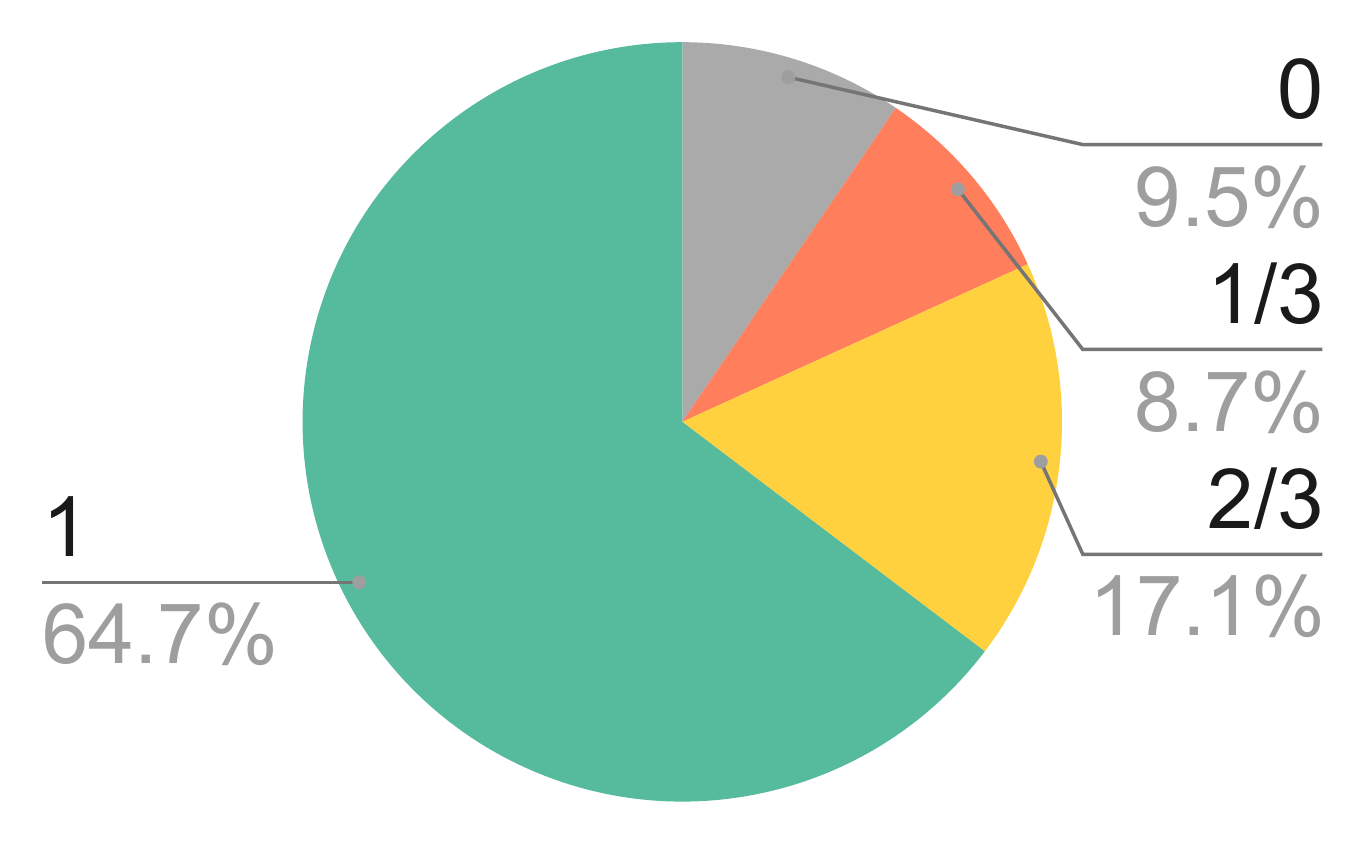}}%
    \caption{\small{Precision Distribution by Query of Logit}}%
    \label{fig:distribution}%
    \vspace{-3mm}
\end{figure}

According to Figure \ref{fig:distribution}, we show the precision distribution of the logit learning-to-rank ranking model.
Over the 250 testing queries, 54.3\% are returned Top-10 with 9 or 10 correct results, and 64.7\% are returned Top-3 with all correct results.
When deploying subjective search systems in production, this allows us to contribute to an existing system 
by rerouting these queries alike.

\section{Conclusion}\label{sec:conclusion}

As user-generated data becomes more prevalent, it plays a critical role when users make decisions about products and services. 
However, existing search systems are not sufficiently supporting search over subjective data like online reviews. 
To bridge this gap, we implemented a production-level subjective search system that achieved 70+\% overall precision and 
90+\% precision on more than half of the benchmark queries.

\smallskip
\noindent
\textbf{Takeaways. }
To build such a production search system, several challenges need to be appropriately addressed. First, the search quality 
of a single search algorithm (e.g., \system, BM25, or BERT) is not enough for production use. 
Through error analysis, we found those models complement each other to some degree. 
We achieved the final good precision by combining these algorithms under a learning-to-rank framework. 
Second, it is nontrivial to evaluate the quality of subjective search since a benchmark of subjective queries does not exist. 
\hl{We constructed a set of over 600 subjective queries from real dialogues, interviews and review tagline summaries.}
We also crowdsourced the query-hotel relevancy labels iteratively and applied careful quality control to maximize the values of the collected labels.
Last but not least, certain engineering efforts are needed for localization when building the search system. 

\hl{As for future work, we plan to deploy the system to other popular production scenarios, such as proposing recommendations when users are seeking for suggestions in chatbots.}
We are also expending the subjective search system to other domains beyond hospitality including housing and restaurant recommendation.

\bibliographystyle{abbrv}
\bibliography{paper}

\end{document}